\documentclass[aps,prb,reprint,superscriptaddress,floatfix]{revtex4-2}

\usepackage{amsmath,amssymb,bm,graphicx}
\usepackage[hidelinks]{hyperref}
\newcommand{\ee}{\mathrm{e}}
\newcommand{\ii}{\mathrm{i}}

\begin{document}

\title{Thermal quasi-Devil's staircase in an anisotropic triangular-lattice Rydberg array}

\author{Jinghao Cao}
\affiliation{Department of Physics, School of Science and Research Center for Industries of the Future, Westlake University, Hangzhou 310030, China}
\affiliation{Institute of Natural Sciences, Westlake Institute for Advanced Study, Hangzhou 310024, China}

\author{Siyi Yang}
\affiliation{Department of Physics, School of Science and Research Center for Industries of the Future, Westlake University, Hangzhou 310030, China}
\affiliation{Institute of Natural Sciences, Westlake Institute for Advanced Study, Hangzhou 310024, China}

\author{Jingya Wang}
\affiliation{Department of Physics, School of Science and Research Center for Industries of the Future, Westlake University, Hangzhou 310030, China}
\affiliation{Institute of Natural Sciences, Westlake Institute for Advanced Study, Hangzhou 310024, China}

\author{Dong-Xu Liu}
\affiliation{Department of Physics, School of Science and Research Center for Industries of the Future, Westlake University, Hangzhou 310030, China}
\affiliation{Institute of Natural Sciences, Westlake Institute for Advanced Study, Hangzhou 310024, China}

\author{Zheng Yan}
\affiliation{Department of Physics, School of Science and Research Center for Industries of the Future, Westlake University, Hangzhou 310030, China}
\affiliation{Institute of Natural Sciences, Westlake Institute for Advanced Study, Hangzhou 310024, China}

\begin{abstract}
A Devil's staircase is a sequence of transitions between topological sectors as a control parameter is varied. Such staircases are predicted in frustrated magnets and lattice gauge theories, but observing them is hard: the winding number that labels each sector is topologically protected, so local quantum dynamics alone cannot move the system from one step to the next. Here we propose and analyze an experimentally feasible way to realize a thermal quasi-Devil's staircase in a finite Rydberg-atom array. We study the anisotropic triangular-lattice Ising antiferromagnet using a dimer mapping, a directed-string description, exact thermodynamics, and Monte Carlo simulations. In the defect-free string manifold, the commensurate--incommensurate onset follows from a simple energy--entropy balance, and the exact Ising critical relation reduces to the same condition when triangle-rule defects are rare. On a finite  lattice, this onset appears as a quasi-Devil's staircase of winding-sector crossovers, which we resolve through winding distributions, structure factors, and directional correlations. The key point is that thermal fluctuations overcome the topological barriers that block sector changes under purely quantum dynamics, and the staircase exists precisely at the finite sizes and tunable effective temperatures of current Rydberg platforms. This makes the effect directly observable in existing experiments. Our comparison also shows that the finite-size steps are not phase transitions: in the thermodynamic limit only a single phase transition survives, and the system crosses over to a defect-dominated paramagnet.
\end{abstract}

\maketitle

\section{Introduction}
\label{sec:introduction}

Programmable Rydberg-atom arrays have established a versatile platform for geometrically frustrated Ising models with tunable geometry and site-resolved readout. 
Pioneering experiments realized transverse-field Ising models on square and triangular arrays with up to 196 atoms~\cite{Scholl2021}, resolved topological spin-liquid correlations via blockade constraints~\cite{Semeghini2021,Verresen2021}, and extended to 256-atom systems with long-range interactions~\cite{Ebadi2021,Chen2023}. 
These advances demonstrate that local defects, strings, winding sectors, and real-space correlations can be reconstructed from repeated snapshots, making triangular Ising benchmarks directly relevant to current experiments.

In the small-field limit (vanishing Rabi coupling), the Rydberg Hamiltonian reduces to a purely Ising model with antiferromagnetic couplings set by the van der Waals potential, with geometric distortions providing a natural route to introduce bond anisotropy. 
Theory predicts stripe and density-wave states, order-by-disorder, and Kosterlitz--Thouless behavior in triangular arrays~\cite{LiRydberg2022,Guo2023}, while related physics appears in the quantum magnet ~\cite{LiTMGO2020} and in artificial dipolar Ising systems~\cite{SmeraldMila2018}. 
The platform has since grown dramatically: thousand-qubit systems are now routine, with roadmaps envisioning up to 10,000 atoms~\cite{Bai2026ManyBody,Wang2026NeutralAtom}; Rydberg simulators have also achieved quantitative ``one-to-one'' simulations of real materials such as TmMgGaO$_4$~\cite{leclerc2026one}. 
Beyond equilibrium phases, novel states such as the quantum floating phase have been observed, directly probing commensurate--incommensurate transitions~\cite{Zhang2025Floating}. 
The Rydberg blockade has been harnessed to implement quantum dimer models on square and triangular geometries, revealing U(1) and $\mathbb Z_2$ spin-liquid ground states~\cite{Zeng2025Dimer,yan2022triangular,yan2023quantum,Zhou2025}, with extensions to three-dimensional spin ice~\cite{Shah2025SpinIce,wang2025doped}. 
Non-equilibrium studies have further revealed curvature-driven domain-wall coarsening across Ising quantum critical points~\cite{manovitz2025quantum,peng2026short}.

The frustrated triangular-lattice Ising antiferromagnet maps to a fully packed dimer model on the dual honeycomb lattice~\cite{R.Moessner2001,Wannier1950,Houtappel1950,Isakov2003,liang2026frustrated,Zhou2023,yan2021widely,yan2021topological}, exposing winding sectors that are precisely the topological sectors of an emergent lattice gauge theory~\cite{moessner2010quantum,yan2022global,yan2023quantum}. 
In the isotropic limit, spin correlations are algebraic and admit a height-model description~\cite{Stephenson1970,BloteHilhorst1982,Alet2005,Alet2006,yan2022height,feng2026thermalization}. 
A small anisotropy selects a stripe reference state; strings relative to it carry positive energy but extensive entropy, yielding a commensurate--incommensurate instability~\cite{PokrovskyTalapov1979,Zhang2016}. 
Closely related quantum triangular-lattice models connect domain-wall density to an incommensurate ordering wave vector~\cite{Zhang2016,Zhou2023,Zhou2025,zhou2022quantum}.

Quantum fluctuations can in principle induce transitions between these winding sectors, giving rise to a Devil's staircase in which the system sequentially populates discrete winding sectors as a control parameter is varied~\cite{papanikolaou2007devil,schlittler2015phase,lan2015emergent}. 
In the thermodynamic limit, an infinite sequence of such transitions can emerge within a finite parameter window. 
The hallmark of this effect is the gradual, stage-like emergence of winding sectors in response to tuning of quantum fluctuations, analogous to climbing a staircase. In a finite-size system, this ideal fractal structure is truncated: only a finite number of discrete steps are resolvable, and the infinite hierarchy of transitions collapses into a finite sequence. We refer to this finite-size counterpart as a \emph{quasi-Devil's staircase}.
However, because the winding number is a topological invariant robust against local perturbations, changing winding sectors in realistic experiments poses a significant challenge: the experimentally prepared state often remains trapped in its initial sector even when the ground state has already transitioned to a different one.

In this work, we propose an experimentally feasible scheme in which thermal fluctuations drive a quasi-Devil's-staircase sequence of winding sectors in a finite-size Rydberg array. 
Both numerical and analytical results demonstrate that local thermal fluctuations are sufficient to overcome the topological barriers between sectors, enabling sequential sector changes that are otherwise difficult to access under purely quantum dynamics. Crucially, this quasi-Devil's staircase is a finite-size effect that vanishes in the thermodynamic limit, where the discrete sectors merge into a single continuous onset. This makes finite-size cold-atom arrays---with their tunable lattice size and effective temperature---the natural platform for observing it.
The scheme is enabled by the ability to control the effective temperature of the Rydberg system, achievable via energy-scale rescaling (tunable by interatomic spacing $R$)~\cite{Lin2026Thermodynamic}, cavity-induced prethermalization~\cite{Mikheev2025Prethermalization}, and accurate thermometry protocols~\cite{Fitzner2026Thermometry}. 
These tools allow systematic exploration of thermodynamic phenomena across a wide effective-temperature range, making the predicted staircase effect directly observable in current-generation Rydberg platforms.

\section{Model and constrained manifold}
\label{sec:model}

\subsection{Anisotropic triangular-lattice Ising model}
We consider a triangular lattice with $N = L_x \times L_y$ sites, where $L_x$ and $L_y$ are the number of sites along the two primitive lattice directions. For the symmetric lattices studied below, $L_x = L_y \equiv L$, and $L$ also denotes the linear system size used in the string counting. Periodic boundary conditions are imposed along both directions unless stated otherwise. In the numerical simulations, $N = L^2$ is the total number of spins.

\begin{figure}[t]
  \centering
  \includegraphics[width=0.98\columnwidth]{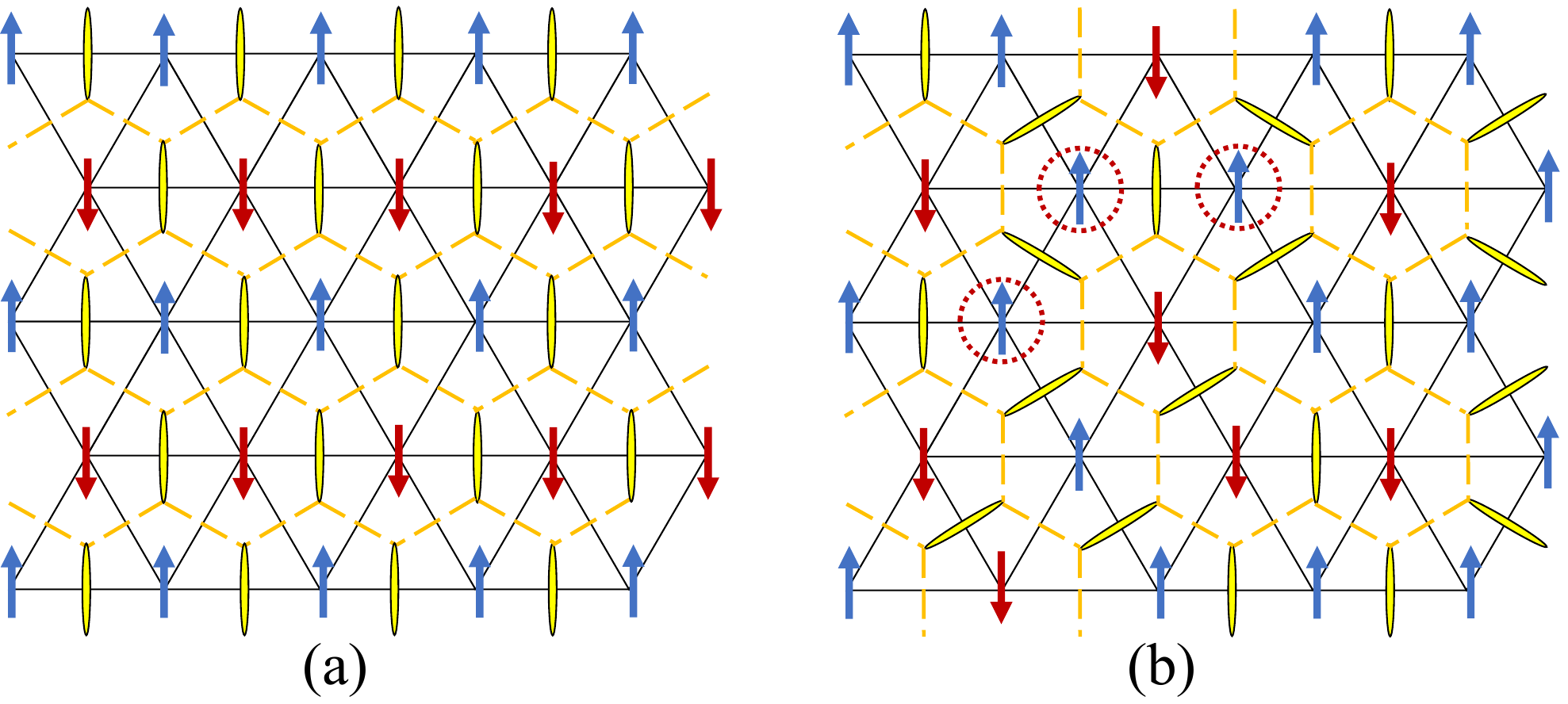}
  \caption{Low-energy spin-to-dimer mapping. 
  Yellow dimers occupy dual edges that cross parallel-spin, and hence frustrated, bonds; dashed lines show the dual honeycomb lattice. 
  (a) The stripe reference state and its staggered dimer covering. 
  (b) A different triangle-rule configuration. 
  The red circles mark flippable spins, for which the associated dual hexagon has three alternating dimers. 
  Flipping a nonflippable spin creates triangle-rule defects.}
  \label{fig:dimer-mapping}
\end{figure}

We focus on thermal dynamics and therefore take the small-field limit of the Rydberg Ising Hamiltonian. The remaining variables are $\sigma_i=\pm1$ on the sites of a triangular lattice, with Hamiltonian

\begin{equation}
H
=J_x\sum_{\langle ij\rangle_x}\sigma_i\sigma_j
+J\sum_{\langle ij\rangle_{\wedge}}\sigma_i\sigma_j ,
\label{eq:model}
\end{equation}
where $J,J_x>0$. The first sum runs over one bond direction, chosen as the $x$ direction, and the second sum runs over the other two directions. We take $J_x<J$. This anisotropy selects a stripe ground-state in which the frustrated bonds lie along the weaker direction, as shown in Fig.\ref{fig:dimer-mapping} (a). Because the van der Waals interaction falls as $r^{-6}$, we retain only nearest-neighbor couplings.


Antiferromagnetic bonds cannot all be satisfied on a triangle. The minimum energy of an isolated triangle is obtained when two bonds join antiparallel spins and the remaining bond joins parallel spins. We call this local condition the \emph{triangle rule}; its unique parallel-spin bond is the frustrated bond. At temperatures well below the cost of violating this rule, the many-spin problem is restricted to configurations with exactly one frustrated bond on every triangle.

Fig.~\ref{fig:dimer-mapping} gives the geometric representation used below. Each triangular plaquette becomes a vertex of the dual honeycomb lattice, and we place a dimer on the dual edge that crosses its frustrated bond~\cite{Wannier1950,Houtappel1950}. Because a bond is shared by two adjacent triangles, one dimer accounts for the frustrated bond of both triangles. The triangle rule therefore requires exactly one dimer at every dual vertex. The low-energy spin manifold is thus mapped to close-packed honeycomb dimers, or equivalently to perfect matchings. Panel (a) shows the stripe state and its staggered dimer covering, while panel (b) shows another covering in the same constrained manifold. A global spin reversal leaves all frustrated bonds unchanged, so the map is two-to-one on a simply connected domain. On a torus, a dimer covering reconstructs periodic spins only when its net winding around the two periodic directions is compatible with the chosen spin boundary conditions.

A local spin flip does not automatically preserve the triangle rule. The six bonds incident on a spin form the six edges of the corresponding dual honeycomb hexagon. Flipping the spin interchanges the satisfied and frustrated status of all six bonds. The final state still has one frustrated bond per triangle only when the hexagon initially contains three dimers on alternating edges. Such a hexagon is \emph{flippable}: the spin flip exchanges its two alternating three-dimer patterns. If the occupied edges do not alternate, the same spin flip creates adjacent triangles with zero or more than one frustrated bond and leaves the constrained manifold. The red circles in Fig.~\ref{fig:dimer-mapping}(b) identify representative flippable spins.

\subsection{Strings and winding sectors}

\begin{figure*}[t]
  \centering
  \includegraphics[width=0.95\textwidth]{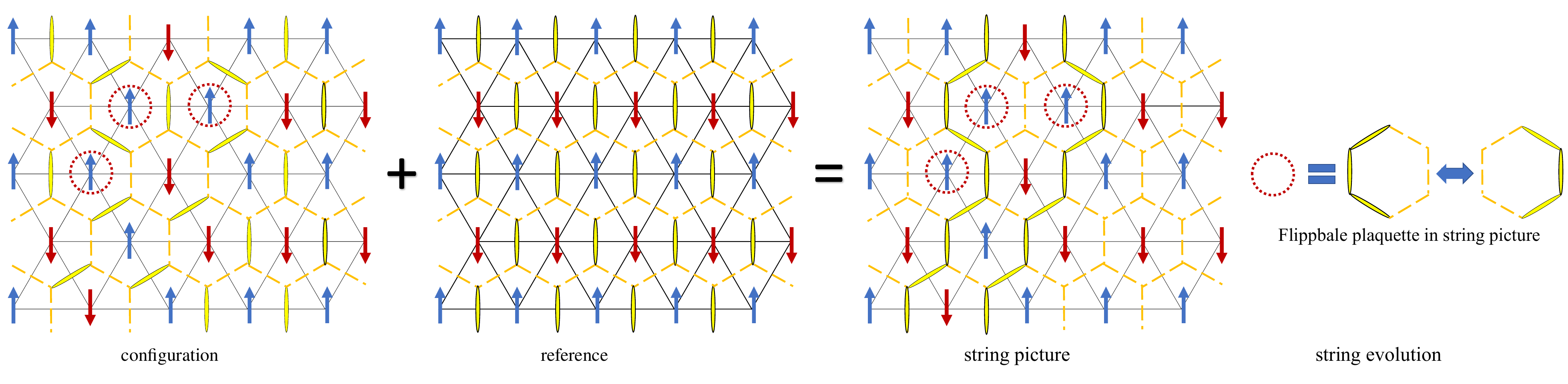}
  \caption{Construction and local motion of strings. 
  The plus sign denotes XOR (the symmetric difference), which for binary dimer occupations is equivalent to addition modulo two. Common dimers cancel, and the remaining edges form
  nonbranching loops. 
  A noncontractible loop is shown as a system-spanning string; the plaquette
  sketches show local deformations that preserve winding.}
  \label{fig:string-construction}
\end{figure*}

Strings are defined relative to the stripe reference rather than inserted as additional objects. 
In Fig.~\ref{fig:string-construction}, the plus sign denotes XOR, or the symmetric difference, of the two dimer coverings. 
For the binary edge occupations $n_e,n_e^{\rm ref}\in\{0,1\}$, XOR is mathematically equivalent to $s_e=(n_e+n_e^{\rm ref})\bmod 2$. 
Thus common dimers cancel, whereas edges occupied in only one covering remain. At each dual vertex, the two coverings either select the same edge, leaving degree zero after cancellation, or select two different edges, leaving degree two.
Consequently, the surviving edges form disjoint, nonbranching loops. 
A flippable move changes the dimers around one dual hexagon. In the string
picture, it replaces one short piece of a loop by another nearby piece. The
loop is therefore deformed only locally, while its winding remains unchanged.
All of these paths are closed on a torus. 
A loop is \emph{contractible} if it can be continuously shrunk to a point
without cutting it. A loop that winds around a periodic direction is
\emph{noncontractible} and cannot be shrunk in this way. When the torus is
drawn as a rectangle, such a loop appears as a system-spanning path, which we
call a \emph{string}~\cite{Kasteleyn1961,Fisher1961,Motruk2012Bose}.

The winding sector records this global wrapping. Choose two noncontractible reference cuts, i.e., closed curves that wrap once around each periodic direction, and count the oriented net number of string crossings through each cut. The resulting pair $(W_x,W_y)$ specifies the homology class of the loop configuration~\cite{moessner2010quantum,yan2022global}. Two configurations belong to the same \emph{winding sector} when they have the same pair, even if their strings differ by arbitrary local deformations. The anisotropy singles out one string orientation, so below we report only the corresponding scalar estimator $W$ and its density $W/L$. With the cut convention used in the simulations, adjacent admissible string sectors differ by $\Delta W=2$, because periodic spin boundary conditions require the total number of string crossings to be even; $N_s$ denotes the analytic number of system-spanning strings, whereas $W$ is the associated XOR-crossing estimator.

The right side of Fig.~\ref{fig:string-construction} also explains why the winding is topological. A flippable plaquette move changes a loop only within a contractible neighborhood and cannot alter its net cut crossings. Thus all local, triangle-rule-preserving moves conserve $(W_x,W_y)$. Thermal spin updates can nevertheless change sectors by leaving the constrained manifold: a nonflippable spin flip creates a pair of adjacent triangles each violating the triangle rule, which are the endpoints of an open string segment; one endpoint can move around a periodic direction, and annihilation of the pair then restores the triangle rule in a different winding sector \cite{yan2023quantum}. This defect-mediated process is the physical origin of thermally activated sector changes in the numerical data.

Within the constrained manifold, the anisotropy supplies a string tension. A segment of string shifts the frustrated bond from the weak-bond stripe pattern to one of the stronger bond directions. Compared to the stripe reference, the corresponding bond-energy change is $2(J-J_x)$ per longitudinal layer. The tension of the string is therefore $J_s=2(J-J_x)$. For $N_s$ system-spanning strings of length $L_y$,

\begin{equation}
E_{N_s}-E_0=N_sJ_sL_y,
\qquad
J_s=2(J-J_x).
\label{eq:string-energy}
\end{equation}

The stripe state has $N_s=0$.  For fixed $N_s$, let $\Omega_{N_s}$ denote the number of allowed noncrossing configurations.  
With $\beta=(k_{\mathrm B}T)^{-1}$, the restricted partition function and free energy are

\begin{align}
Z_{N_s}^{(0)}
&=\Omega_{N_s}\exp(-\beta N_sJ_sL_y), \\
F_{N_s}^{(0)}
&=N_sJ_sL_y-k_{\mathrm B}T\ln\Omega_{N_s}.
\label{eq:sector-free-energy}
\end{align}

\begin{figure}[t]
  \centering
  \includegraphics[width=1\columnwidth]{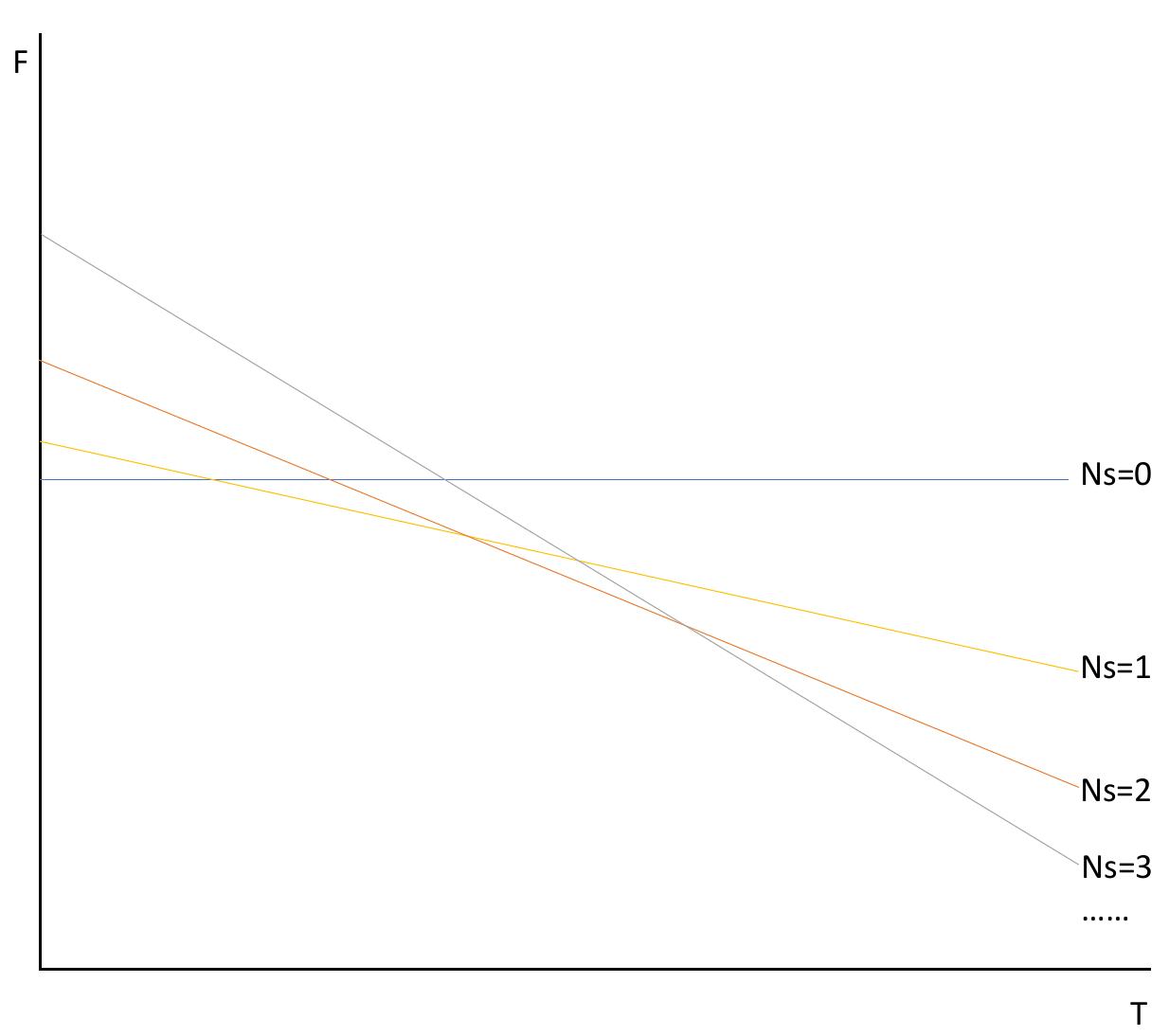}
  \caption{Schematic, restricted free energies for several finite-size winding sectors. The intercept of each line is the sector energy and its slope is minus the sector entropy. Crossings illustrate finite-size sector selection; the lower envelope is the equilibrium restricted free energy. }
  \label{fig:sector-free-energy}
\end{figure}

The superscript $(0)$ emphasizes that each triangle satisfies the triangle rule. Fig.~\ref{fig:sector-free-energy} visualizes Eq.~\eqref{eq:sector-free-energy}. 
Each line represents a finite-size sector with different numbers of strings $N_s$: its vertical intercept is the sector energy, and its slope is minus the sector entropy. 
The equilibrium restricted free energy is the lower envelope of these lines. 
Their crossings describe finite-size sector selection and are not separate bulk phase transitions.

Fig.~\ref{fig:sector-free-energy} shows a clear physical picture that the winding sectors will appear one by one while the temperature increases for a finite size array. 
The fundamental reason is that more strings bring more entropy at higher temperatures, because flippable plaquettes obeying the triangle rule can only be located at the corners of a string [see Fig.\ref{fig:string-construction}].
This sequence of sector crossings is the finite-size mechanism behind the quasi-Devil's staircase discussed in Sec.~\ref{sec:string-solution}.

\section{Exact onset in the string manifold}
\label{sec:string-solution}

\subsection{Single-string counting}

\begin{figure}[t]
  \centering
  \includegraphics[width=0.92\columnwidth]{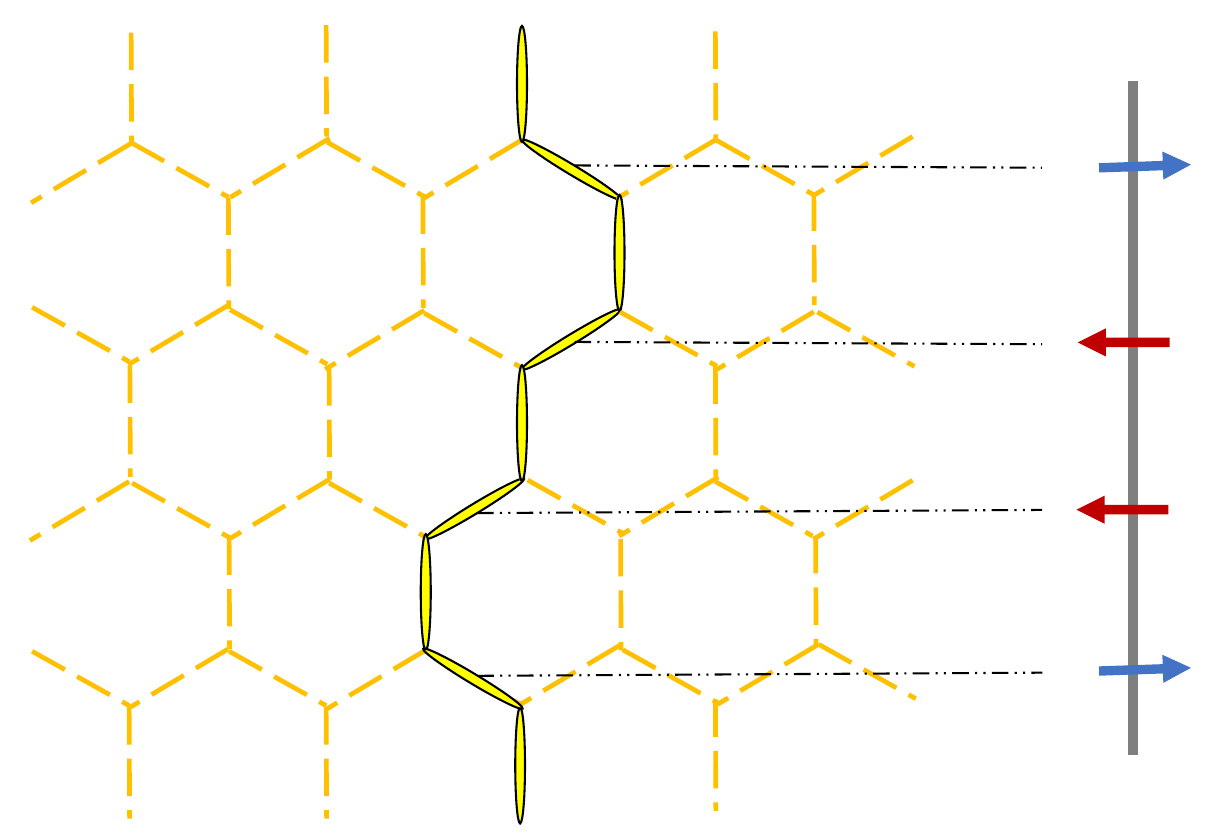}
  \caption{Geometry used for single-string counting. The left panel shows a
  directed string crossing successive longitudinal layers. The dashed
  horizontal lines mark these layers; the blue and red arrows on the right
  show the two allowed transverse shifts at one layer.}
  \label{fig:string-chain}
\end{figure}

Fig.~\ref{fig:string-chain} shows how we count the paths of one
system-spanning string. The string crosses $L$ longitudinal layers. At each
layer it can take one of the two transverse steps shown by the blue and red
arrows. We record the two choices by
$\tau_\ell=+1$ and $\tau_\ell=-1$, respectively. Thus
$(\tau_1,\tau_2,\ldots,\tau_L)$ is a one-dimensional binary sequence that
records the shape of the two-dimensional string; $\tau_\ell$ is not an
original spin $\sigma_i$~\cite{Zhang2013Chiral}.

The string can start on any of the $L$ transverse rows. Periodic boundary
conditions require it to return to its starting row, so the total transverse
displacement must vanish:
\begin{equation}
\sum_{\ell=1}^{L}\tau_\ell=0.
\label{eq:string-closure}
\end{equation}

For even $L$, this means that $L/2$ steps point in each transverse direction.
The number of closed one-string paths is therefore

\begin{equation}
\Omega_1=L\binom{L}{L/2}.
\label{eq:one-string-count}
\end{equation}

This is the standard directed-string counting used in the
Kasteleyn--Pokrovsky--Talapov description of weighted dimers
~\cite{Kasteleyn1961, PokrovskyTalapov1979}.

We measure the free energy relative to the stripe state, which contains no
system-spanning string. Adding one string costs the energy $J_sL$, while its
$\Omega_1$ possible paths contribute the entropy
$k_{\mathrm B}\ln\Omega_1$. Thus

\begin{align}
\Delta F_1^{(0)}
&=J_sL-k_{\mathrm B}T
\ln\!\left[L\binom{L}{L/2}\right].
\label{eq:one-string-free-energy}
\end{align}

Using the central-binomial asymptotic form~\cite{Elezovic2014Asymptotic}

\begin{equation}
\binom{L}{L/2}\sim \sqrt{\frac{2}{\pi L}}\,2^L
\qquad (L\to\infty).
\label{eq:central-binomial-asymptotic}
\end{equation}

the entropy per layer becomes

\begin{equation}
\frac{1}{L}\ln\Omega_1
=\ln2+O\!\left(\frac{\ln L}{L}\right).
\label{eq:string-entropy-asymptotic}
\end{equation}

The factor $2^L$ is the leading number of unconstrained step sequences. The
closure condition and the choice of starting row give only finite-size
corrections. Therefore,

\begin{equation}
\frac{\Delta F_1^{(0)}}{L}
=J_s-k_{\mathrm B}T\ln2+O\!\left(\frac{\ln L}{L}\right).
\label{eq:string-free-energy-density}
\end{equation}

At low temperature, the energy cost is larger than the entropy gain, so a
string is unfavorable. The string onset $T_s$ is defined by the temperature at
which the free-energy cost per unit length vanishes:

\begin{equation}
k_{\mathrm B}T_s=\frac{J_s}{\ln2}
=\frac{2(J-J_x)}{\ln2}.
\label{eq:string-critical-temperature}
\end{equation}

This is the zero-density string onset predicted by the defect-free string counting model. It is not a thermodynamic phase transition; the true critical point $T_c$ of the full Ising model is discussed in Sec.~\ref{sec:exact}.

\subsection{Finite-size sector crossings}

Let $\Omega_n$ denote the number of allowed configurations containing $n$ system-spanning strings. 
In general, the multi-string count is not $(\Omega_1)^n$ since strings are indistinguishable, cannot cross, and reduce one another's available transverse phase space~\cite{Wang2026Quantum}.
Therefore, exact finite-size values require a non-crossing-path determinant, a path transfer matrix, or the equivalent free-fermion construction.

For finite $L_x$ and $L_y$, adding the $n$th string changes the energy by $J_s L_y$ and the entropy by $k_B \ln(\Omega_n/\Omega_{n-1})$. 
The two sectors have equal statistical weight when their free energies are equal, $F_n^{(0)}=F_{n-1}^{(0)}$:

\begin{equation}
k_{\mathrm B}T_{n-1\rightarrow n}
=\frac{J_sL_y}{\ln(\Omega_n/\Omega_{n-1})}.
\label{eq:sector-crossing}
\end{equation}

This equation gives the temperature at which two adjacent restricted sectors have equal statistical weight.
On a finite lattice, sectors with different string numbers cross at different temperatures, so the system passes through them one by one as the temperature increases.
If the transverse size is made larger while the string numbers are kept fixed, these crossing temperatures move closer together and, in the thermodynamic limit, all collapse onto the same onset scale $T_s$.
This does not mean that the string density jumps to its maximum at $T_s$.
Rather, the onset at $T_s$ describes only the limit in which the string number is held fixed while the system size grows, so that the string density tends to zero. 
If the string density is instead held fixed as the system size grows, the temperature at which that density becomes favorable does not approach $T_s$; it remains higher than $T_s$ and depends on the density. Thus $T_s$ marks the point where a dilute gas of strings first becomes favorable, not the point at which the system reaches any finite string density.

This size dependence has a direct experimental consequence. The staircase is not a thermodynamic-limit phenomenon; it exists only at finite size, where the winding sectors remain discrete. Real cold-atom arrays are inherently finite, so they naturally realize the regime in which the staircase is visible, and the number of resolved steps can be tuned by changing the lattice size.

\section{Exact thermodynamic benchmark}
\label{sec:exact}

The string calculation keeps only configurations in which every triangle has exactly one frustrated bond. 
The full Ising model also contains higher-energy configurations that violate this local condition. 
For example, a triangle with all three spins parallel has three frustrated bonds. 
The exact solution includes these defect-containing configurations together with the low-energy configurations~\cite{Houtappel1950,Onsager1944,Kaufman1949,Schultz1964}.  

For the conventional exact-solution notation, one bond energy is written as $- \mathcal{J}_a \sigma_i \sigma_j$.
With signed $K_a=\beta\mathcal J_a$, the exact free energy per spin can be written

\begin{equation}
f=-k_{\mathrm B}T\left[
\ln2+\frac{1}{8\pi^2}
\int_0^{2\pi}dw_1\int_0^{2\pi}dw_2\,\ln A(w_1,w_2)
\right],
\label{eq:exact-free-energy}
\end{equation}

where

\begin{align}
A={}&\prod_{a=1}^3\cosh(2K_a)
+\prod_{a=1}^3\sinh(2K_a)\notag\\
&-\sinh(2K_1)\cos w_1
-\sinh(2K_2)\cos w_2\notag\\
&-\sinh(2K_3)\cos(w_1+w_2).
\label{eq:free-energy-kernel}
\end{align}

The variables $w_1$ and $w_2$ are the reciprocal-space phases generated by translations along the two primitive lattice vectors. 
The function $A(w_1,w_2)$ is the spectral determinant of the quadratic mode obtained in the exact transfer-matrix, or equivalently Pfaffian solution. 
Its logarithm sums the free-energy contribution of all momentum modes. 
As long as $A$ remains strictly positive, the momentum integral is analytic. 
When its minimum reaches zero, a singularity appears in the free energy and this singularity marks the critical point.

For two equal strong antiferromagnetic bond magnitudes
$|\mathcal J_1|=|\mathcal J_2|\equiv J$ and a weaker magnitude
$|\mathcal J_3|=J_x<J$, the exact critical temperature $T_c$ satisfies

\begin{equation}
k_{\mathrm B}T_c=
\frac{2(J-J_x)}
{\ln\!\left[1+\sqrt{1+\exp(-4J_x/k_{\mathrm B}T_c)}\right]}.
\label{eq:exact-critical-temperature}
\end{equation}

It follows that

\begin{equation}
\frac{2(J-J_x)}{\ln(1+\sqrt2)}
<k_{\mathrm B}T_c<
\frac{2(J-J_x)}{\ln2}.
\label{eq:critical-bounds}
\end{equation}

When $4J_x/(k_{\mathrm B}T_c)\gg1$, triangle-rule defects are exponentially suppressed and

\begin{equation}
k_{\mathrm B}T_c
\simeq \frac{2(J-J_x)}{\ln2}
=k_{\mathrm B}T_s.
\label{eq:exact-string-limit}
\end{equation}

The exponential term in Eq.~\eqref{eq:exact-critical-temperature} is the leading effect omitted from the defect-free string partition function.
Thus the elementary counting argument reproduces a controlled limit of the complete Ising solution, rather than an independent thermodynamic transition. In other words, $T_s$ is not an independent transition; it is the $J_x/(k_{\mathrm B}T_c)\to\infty$ limit of the exact $T_c$.

\section{Numerical diagnostics}
\label{sec:numerics}

\subsection{Protocol and observables}

We generate the numerical data with single-spin Metropolis updates~\cite{Metropolis1953} and periodic boundaries.  One sweep contains $L^2$ random single-site attempts.  
The simulated Hamiltonian is

\begin{equation}
H_{\mathrm{MC}}
=0.9\sum_{\langle ij\rangle_x}\sigma_i\sigma_j
+\sum_{\langle ij\rangle_{\wedge}}\sigma_i\sigma_j,
\label{eq:mc-hamiltonian}
\end{equation}

so the simulations and analytic theory use the same couplings.  
Each temperature--bin pair starts from the zero-winding stripe state and is thermalized at its target temperature.  We parallelize the bins with OpenMP and perform reciprocal-space and correlation analyses locally.

For the winding scan, we use $L=48$, $5\times10^4$ thermalization sweeps, $10^5$ measurement sweeps, and 12 independent bins over $0\leq T/J\leq2.00$ in steps of $0.01$.  
Six randomly placed reference cuts are measured in every sweep.  For $S(\bm q)$, we use the same lattice size, sweep counts, and bin number at $T/J=0.20$, $0.35$, $0.50$, $0.75$, and $1.00$.
For directional correlations, we use $L=160$, $10^5$ thermalization sweeps, $10^5$ measurement sweeps, and 30 bins at $T/J=0.20$, $0.40$, $0.60$, $0.80$, $1.00$, and $1.20$.  We sample correlations every 25 sweeps from 64 random origins up to $r=L/2$.

The simulation measures the winding number $W$, not the analytic string count
$N_s$. For each production sweep, the code selects six translated cuts and
computes the XOR crossing count on each cut. For every site on a cut, it checks
the two bonds crossing the cut: a bond contributes one when its frustrated
status differs from that of the stripe reference, and zero otherwise. The
sum over the cut is the integer sample $W$. Translating the cut changes where
a fluctuating loop is crossed, but not its topological sector, so averaging
over the six cuts reduces sampling noise.

The code accumulates these integer samples in a histogram. After all $12$
independent bins and production sweeps are combined, it forms
\begin{equation}
P(W,T)=\frac{N(W,T)}{N_{\mathrm{sweep}}N_{\mathrm{bin}}N_{\mathrm{cut}}},
\qquad \sum_W P(W,T)=1,
\label{eq:winding-distribution}
\end{equation}
where $N(W,T)$ is the number of samples with winding number $W$. The mean
$\langle W\rangle=\sum_W W P(W,T)$ can hide the separate sectors, whereas
$P(W,T)$ displays their individual statistical weights. The analytic $N_s$
counts system-spanning strings in the restricted dimer description; it is not
identified with the measured $W$ without an additional convention-dependent
mapping.

The measured structure factor is

\begin{equation}
S(\bm q)=\frac{1}{N}
\left\langle\left|
\sum_i\sigma_i\ee^{\ii\bm q\cdot\bm r_i}
\right|^2\right\rangle ,
\label{eq:structure-factor}
\end{equation}

and directional correlations are

\begin{equation}
C_\alpha(r)=\frac{1}{N}
\sum_i\langle\sigma_i\sigma_{i+r\hat{\bm e}_\alpha}\rangle .
\label{eq:directional-correlation}
\end{equation}

Here $\bm q$ is the reciprocal-space wave vector, $\bm r_i$ is the position
of site $i$, and $\hat{\bm e}_\alpha$ ($\alpha=x,y$) are the two primitive
lattice directions.
In the anisotropic lattice, $\hat{\bm e}_x$ points along the weaker bond direction and $\hat{\bm e}_y$ along one of the two stronger bond directions.
In both equations, $i$ runs over all $N$ lattice sites. In
$C_\alpha(r)$, the partner of site $i$ is displaced by
$r\hat{\bm e}_\alpha$ and wrapped back into the lattice by the periodic
 boundary conditions.

The factor $1/N$ in $S(\bm q)$ is the standard normalization of the static
structure factor. It does not make the peak height independent of system
size: a perfectly coherent peak has $S(\bm q)\sim N$. For the plots, we apply
one further normalization and use the dimensionless intensive intensity
\begin{equation}
\widetilde S(\bm q)=\frac{S(\bm q)}{N}.
\label{eq:normalized-structure-factor}
\end{equation}
For Ising spins, an ideal coherent peak has $\widetilde S(\bm q)=1$.

A peak in $S(\bm q)$ identifies a periodic spin pattern that is strongly represented in the spin configurations. 
Its position gives the ordering wave vector. We use $\widetilde S(\bm q)$ when
comparing the coherent peak weight across system sizes. 
A narrow peak indicates correlations extending over longer distances, although on a finite lattice the peak width is also limited by the system size. 
By contrast, $C_\alpha(r)$ measures directly whether two spins separated by $r$ along direction $\alpha$ tend to point the same way or opposite ways. 
Its sign carries the modulation phase: a positive value favors parallel alignment and a negative value favors antiparallel alignment. 
Its envelope measures how far the spatial coherence persists before decaying away.

For a periodic lattice, Eq.~\eqref{eq:structure-factor} equals the discrete Fourier transform of the translationally averaged two-point function at the allowed reciprocal vectors.  
In post-processing, we retain the signs of the sublattice-resolved correlations and include the basis phases before the Fourier transform.  
We verify the resulting $S(\bm q)$ against the direct spin-amplitude expression in Eq.~\eqref{eq:structure-factor}.  The largest pointwise discrepancy is $2.84\times10^{-14}$.

\subsection{Winding and reciprocal-space signatures}

\begin{figure*}[t]
  \centering
  \includegraphics[width=0.98\textwidth]{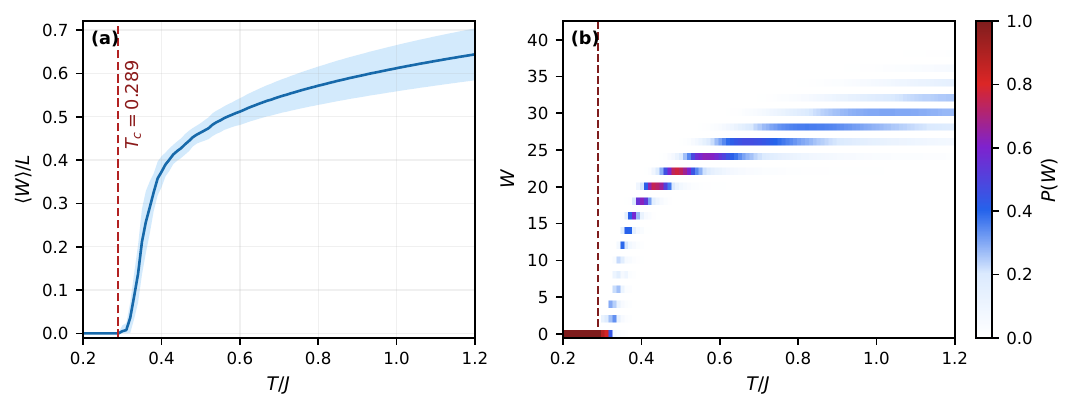}
  \caption{Finite-size winding diagnostics for the anisotropic $L=48$ model with $J=1$ and $J_x=0.9$.  
  (a) Mean winding density; shading gives the root mean square (RMS) width of the sampled winding-sector distribution. 
  (b) Full distribution $P(W,T)$.  
  The dashed line marks the exact critical temperature, $k_{\mathrm B}T_c/J=0.288539$.  
  In the low-defect regime, periodic closure selects even winding sectors.}
  \label{fig:winding}
\end{figure*}

We first test the topological prediction directly. 
Fig.~\ref{fig:winding}(a) shows the mean winding density, which indicates when the ensemble leaves the stripe sector on average. 
However, the mean does not reveal whether it comes from a single broad distribution or from successive occupation of discrete sectors. 
Panel (b) retains the full distribution. A localized maximum in $P(W,T)$ that persists over a range of temperature indicates that the corresponding finite-size winding sector has appreciable statistical weight.

Fig.~\ref{fig:winding} resolves the finite-size sector sequence.  
Below the analytic onset, $P(W,T)$ is concentrated at $W=0$.  
Weight first appears in $W=2$ near $T/J=0.30$ and then shifts successively to higher even sectors.  
Between $T/J\simeq0.32$ and $0.50$, the individual sectors remain distinguishable as separate localized maxima.  
At higher temperature these maxima overlap, and the mean winding density varies smoothly even though the underlying sectors remain discrete. 
The exact critical temperature $T_c$ falls at the onset of the observed increase, but the numerically visible signal is delayed to $T/J \simeq 0.30$ by the finite lattice width, the discrete spacing between sectors, and the activation barrier of local spin updates.

The ridges are finite-size winding-sector crossovers, not separate thermodynamic phases.  Equation~\eqref{eq:sector-crossing} explains why a finite torus resolves neighboring sector free energies one by one.  
As the transverse width increases, the low-density threshold spacing collapses.  
The exact free energy has only the single onset in Eq.~\eqref{eq:exact-critical-temperature}.

The same sector migration has a reciprocal-space signature. 
The position of a structure-factor peak gives the wavelength of the dominant spin modulation. 
As strings proliferate, they insert transverse shifts into the stripe pattern, so the modulation wavelength changes and the original stripe peak splits into symmetry-related peaks that move along the Brillouin-zone boundary. 

\begin{figure*}[t]
  \centering
  \includegraphics[width=0.99\textwidth]{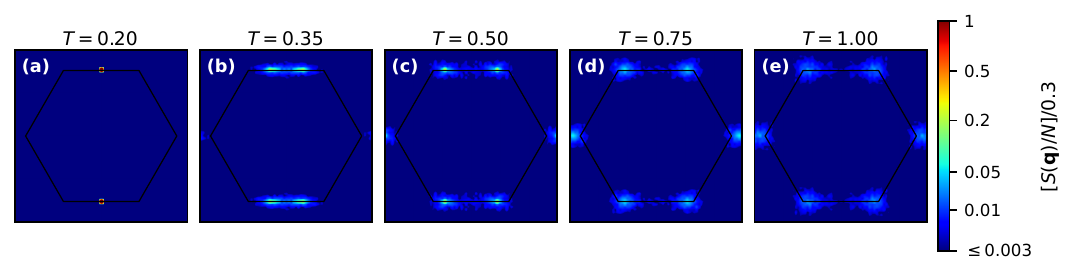}
  \caption{Structure factor of the anisotropic $L=48$ model at $T/J=0.20, 0.35, 0.50, 0.75, 1.00$.
  The black hexagon marks the first Brillouin-zone boundary and does not crop the data.
  Equation~\eqref{eq:structure-factor} defines the standard structure factor
  $S(\bm q)$, whose coherent peak grows with system size. For plotting, we
  further normalize by $N$ and use the dimensionless intensity
  $I=\widetilde S(\bm q)/0.3=[S(\bm q)/N]/0.3$ with the same
  nonlinear color mapping $[(I-0.003)/(1-0.003)]^{0.35}$ above a common
  display floor of $I=0.003$ to make weak split peaks visible.
  Values at or below this floor share the same dark-blue color in all panels.
  The colorbar labels give $I$; values above $I=1$ are clipped.
  The stripe peaks split and move along the zone boundary as winding sectors enter.
  At higher temperature, triangle-rule defects broaden and weaken the peaks.}
  \label{fig:qspace}
\end{figure*}

Fig.~\ref{fig:qspace} follows this motion from the commensurate state in panel (a), through the split peaks in panels (b)--(d), to the defect-dominated response in panel (e). 
At high temperature, triangle-rule defects create many incompatible local patterns, so the peaks become weaker and broader, indicating that the configurations no longer share a coherent modulation over long distances. This high-temperature regime is a defect-dominated paramagnet.

At $T/J=0.20$, the locally stable stripe state produces resolution-limited
boundary peaks.  Just above the onset, the maxima leave the commensurate
stripe wave vector and form symmetry-related incommensurate pairs.  Their
displacement grows between $T/J=0.35$ and $0.75$, in parallel with the dominant
winding sector.  By $T/J=1.00$, the peaks are weak and diffuse.  Thus, the
winding-sector sequence and peak motion diagnose the same change of
modulation, while the peak width records the loss of spatial coherence.

The exact solution supplies the thermodynamic benchmark for the numerical scan.  
In the dilute-string picture, one unit of string length costs $J_s$ and contributes $\ln2$ local path choices.  
Triangle-rule defects modify this balance through the exponential term in Eq.~\eqref{eq:exact-critical-temperature}.  
For the simulated anisotropy,
\begin{equation}
\frac{k_{\mathrm B}T_s}{J}=0.288539008,
\qquad
\frac{k_{\mathrm B}T_c}{J}=0.288538611.
\label{eq:comparison-temperatures}
\end{equation}
The defect correction is below the resolution of the scan.  The numerical onset near $T/J\simeq0.30$ in finite size is therefore consistent with the analytic scale, given the scan spacing and local-dynamics barrier.

The condition $F_n^{(0)}=F_{n-1}^{(0)}$ defines a finite-size crossing in a restricted configuration space.  
Only the lower envelope of all sectors is an equilibrium restricted free energy.  
A phase transition requires a nonanalyticity that survives the thermodynamic limit.  
When triangle-rule defects proliferate at higher temperature, $W$ remains a useful geometric diagnostic but no longer counts close-packed dimer strings literally.

\begin{figure*}[t]
  \centering
  \includegraphics[width=0.8\textwidth]{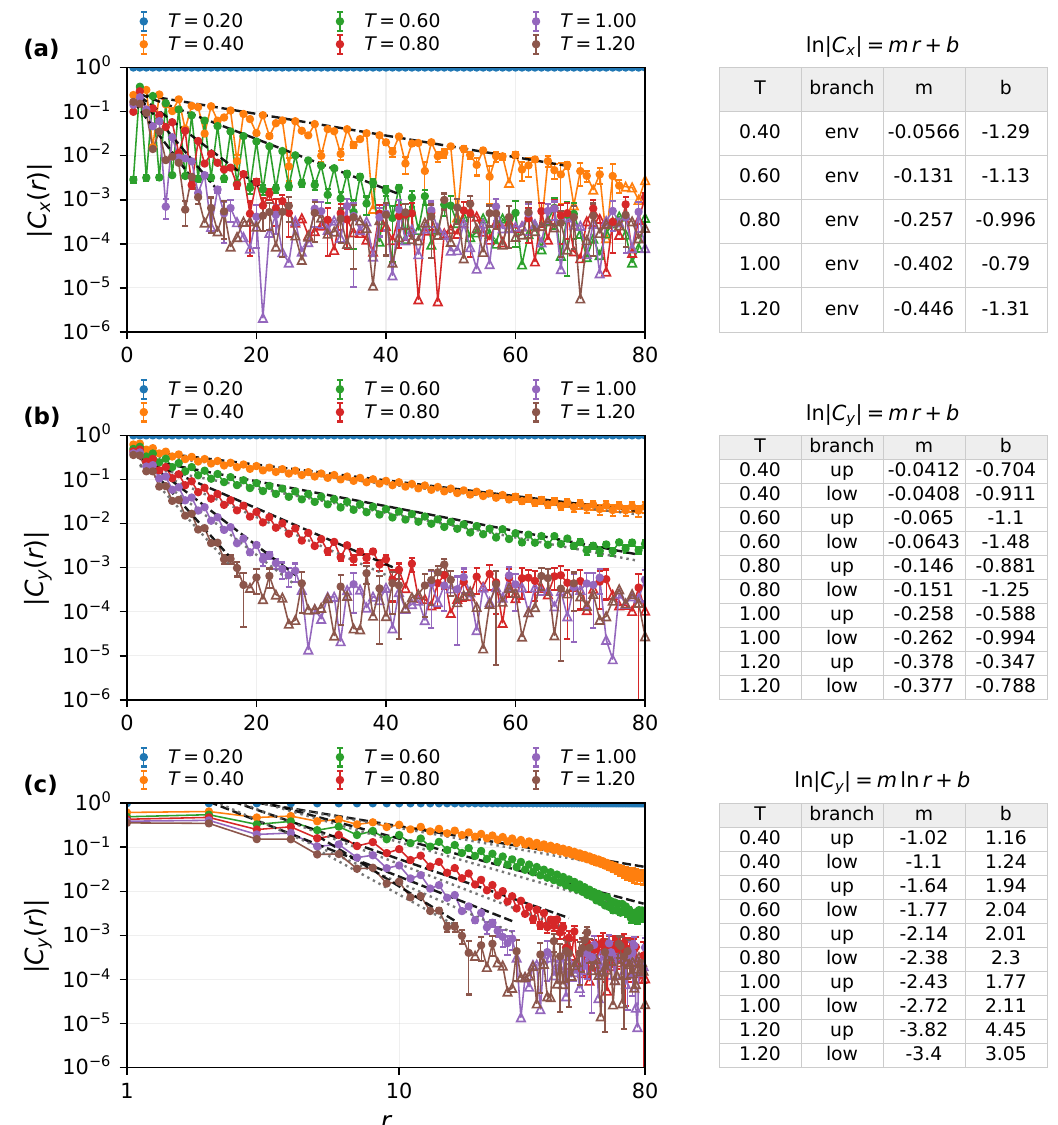}
  \caption{Directional correlations for the anisotropic $L=160$ model, averaged over 30 independent bins. Each row shows the data and fits on the left and fitted parameters on the right.
  (a) Transverse $|C_x(r)|$ on a semilogarithmic scale.
  Exponential fits use only local maxima that pass a three-standard-error threshold; smaller points remain visible but do not enter the fit.  
  (b) Longitudinal $|C_y(r)|$ with exponential fits.
  (c) The same longitudinal data with power-law fits.
  Filled circles show means whose one-standard-error interval excludes zero.
  Where $|C_\alpha(r)|\leq\mathrm{SE}$, an open triangle replaces the circle
  at the measured magnitude, and the entire error bar is omitted.
  Connecting lines are retained for all points. Magnitudes below the
  logarithmic plotting range are displayed at its lower edge.
  Dashed and dotted lines show fits. The tables give slope $m$ and intercept $b$
  for $\ln|C_\alpha|=mr+b$ (exponential) or $\ln|C_\alpha|=m\ln r+b$ (power law).
  Branch labels env, up, and low denote the envelope, upper, and lower branches.}
  \label{fig:correlation-fits}
\end{figure*}

\subsection{Directional correlations}
Fig.~\ref{fig:correlation-fits} shows the same behavior in real space. 
The top row, Fig.~\ref{fig:correlation-fits}(a), plots the transverse correlation $|C_x(r)|$.
Its oscillating branches are the real space trace of the incommensurate modulation that appears as split peaks in Fig.~\ref{fig:qspace}. 
We plot absolute values only to compare decay envelopes; the structure-factor calculation retains signed correlations. 
The middle and bottom rows use the same longitudinal data.
Panel (b) tests exponential decay, and panel (c) tests power-law decay.
So the two fits can be compared directly.

At $T/J = 0.20$, the prepared stripe state retains apparent long-range coherence. 
Above the winding onset, the transverse correlation separates into oscillatory branches whose separation exceeds the error bars, confirming that the splitting reflects incommensurate modulation rather than sampling noise. 
The transverse envelope rapidly reaches the statistical floor, whereas the longitudinal correlation remains measurable at larger separations and decays more slowly. 
Over the restricted longitudinal window near the onset, the exponential fit is visibly better than the power-law fit. 
No nonzero large-distance plateau is resolved above the onset, so the data show no evidence for long-range magnetic order in that regime. 
Together with the split structure-factor peaks, these results demonstrate directional anisotropy and incommensurate transverse modulation.

\section{Conclusion and outlook}
\label{sec:conclusion}
We have shown that the anisotropic triangular-lattice Ising antiferromagnet hosts a thermal quasi-Devil's staircase of topological winding sectors. The dimer mapping converts the constrained spin manifold into close-packed dimers on the dual honeycomb lattice, and the symmetric difference with the stripe reference converts each dimer configuration into noncrossing strings. A single string costs energy $J_s L$ but gains entropy $k_{\mathrm B}L\ln2$, so the zero-density string onset is
$k_{\mathrm B}T_s=2(J-J_x)/\ln2$.
The exact Ising critical relation reduces to the same scale when triangle-rule defects are exponentially rare, establishing that the elementary counting argument captures a controlled limit of the full solution rather than an independent transition. The dilute-string picture and the exact transfer-matrix solution therefore describe the same physics from complementary directions.

The finite-size consequences of this picture are directly observable. On a periodic lattice, the discrete winding sectors cross one by one as temperature increases, producing a quasi-Devil's staircase that merges into a single continuous onset in the thermodynamic limit. For $J=1$ and $J_x=0.9$, Monte Carlo simulations resolve this sequence of even winding sectors, and three independent observables diagnose the same crossover: the winding distribution $P(W,T)$ tracks the topological sector population, the structure factor $S(\bm q)$ develops incommensurate peaks that move along the Brillouin-zone boundary, and the directional correlations $C_\alpha(r)$ reveal anisotropic decay with oscillatory transverse branches but no long-range plateau. The qualitative agreement among these observables, and their consistency with the analytic onset scale, confirms that the staircase is an entropy-driven incommensurate phenomenon rather than a numerical artifact.

Two features of this result are particularly relevant for programmable quantum simulators. First, the staircase exists only at finite size: it is a truncation of the ideal fractal structure, and the number of resolved steps is set by the lattice dimensions. This makes finite cold-atom arrays---which operate with tunable lattice size and effective temperature---the natural platform for observing it. Second, the topological protection of the winding number, which suppresses sector changes under purely quantum dynamics, is overcome by thermal fluctuations through defect-mediated processes. This suggests a general strategy for exploring topological sector selection in constrained quantum matter: rather than driving the system across a quantum phase transition, one can exploit thermal activation at experimentally accessible effective temperatures. Extending this idea to other constrained lattices, to quantum regimes where coherent dynamics competes with thermal activation, and to nonequilibrium protocols where the staircase structure may leave a dynamical signature, are natural directions for future work. More broadly, the connection between dimers, strings, and exact Ising thermodynamics provides a compact framework for interpreting constrained statistical mechanics across artificial spin systems, frustrated magnets, and optimization problems with local compatibility rules.

\begin{acknowledgements}
We thank Zheng Zhou, Xu Zhang, Yang Qi, Zi Yang Meng and Xue-Feng Zhang for helpful discussions, especially Zheng Zhou and Xu Zhang for guidance on the analytical solutions.
This project is supported by the Scientific
Research Project (No. WU2025B011), Feng-Ying Career Development Chair Fund and the Start-up Funding of Westlake
University, and the Natural Science Foundation of China (Grant No. 12674188). The authors thank the IT service office and the high-performance
computing center of Westlake University.
\end{acknowledgements}

\bibliographystyle{apsrev4-2}
\bibliography{incmmsrt}

\end{document}